\documentclass[journal]{IEEEtran}

\usepackage{cite}
\usepackage{t1enc}
\usepackage[cmex10]{amsmath}
\usepackage{amssymb}
\usepackage{array}
\usepackage{mdwmath}
\usepackage{mdwtab}
\usepackage[dvipsnames]{xcolor}
\usepackage{pgfplots}

\usetikzlibrary{plotmarks}
\pgfplotsset{compat=newest}
\pgfplotsset{plot coordinates/math parser=false}
\newlength\figureheight
\newlength\figurewidth 
\PassOptionsToPackage{pdf}{pstricks,pst-node,pst-plot}
\usepackage{pstricks,pst-node,pst-plot}
\usepackage{graphicx}
\usepackage{amsmath,amssymb,array,multirow,bigdelim,dsfont}
\usepackage[amssymb]{SIunits} 
\usepackage{url}
\usepackage{t1enc}
\usepackage{url}   
\usepackage{booktabs}
\usepackage{array}
\usepackage[draft]{fixme}

\begin{document}
\title{Modelling of Path Arrival Rate for In-Room Radio Channels with
  Directive Antennas}
\author{Troels~Pedersen\thanks{\today,  This work is supported 
by  the  Cooperative  Research  Project
VIRTUOSO,  funded  by  Intel  Mobile
Communications, Keysight, Telenor, Aalborg University, and the Danish National
Advanced Technology Foundation. T.  Pedersen  is  with  the  Department   of  Electronic
Systems,  Section  Wireless  Communication  Networks,  Aalborg  University,
Aalborg, 9220, Denmark (e-mail:  troels@es.aau.dk). } }

\maketitle

 \begin{abstract}
   We analyze the path arrival rate for an inroom radio channel with
   directive antennas. The impulse response of this channel exhibits a
   transition from early separate components followed by a diffuse
   reverberation tail. Under the assumption that the transmitter's (or
   receiver's) position and orientation are picked uniformly at random
   we derive an exact expression of the mean arrival rate for a
   rectangular room predicted by the mirror source theory.  The rate
   is quadratic in delay, inversely proportional to the room
   volume, and proportional to the product of beam coverage fractions
   of the transmitter and receiver antennas. Making use of the exact
   formula, we characterize the onset of the
   diffuse tail by defining a ``mixing time'' as the point in time
   where the arrival rate exceeds one component per transmit pulse
   duration.  We also give an
   approximation for the power-delay spectrum. It turns out that the
   power-delay spectrum is     unaffected by the antenna
   directivity. However,  Monte Carlo simulations show that antenna
   directivity does  indeed play an important role for the distribution of
   instantaneous mean delay and rms delay spread.  \end{abstract}

\IEEEpeerreviewmaketitle

\section{Introduction}
\label{sec:introduction}
Stochastic models for the channel impulse response are useful tools
for the design, analysis and simulation of systems for radio localization
and communications. These models allow for tests via
Monte Carlo simulation and in many cases provide analytical results
useful for system design.  Many such models exist for the complex
baseband representation of the signal at the receiver
antenna \footnote{Here we omit any additive terms due to noise or 
interference  as our focus is on characterizing the contribution 
related to the transmitted signal.},   
\begin{equation}
  \label{eq:66}
  y(\tau) = \sum_{k }\alpha_k s(\tau-\tau_k),
\end{equation}
where term $k$ has delay $\tau_k$ and complex gain $\alpha_k$ and
$s(\tau)$ is the complex baseband representation of the transmitted
signal \cite{Hashemi1993}. These gains and delays form a marked  point process with points
$\{\tau_k\}$ and marks $\{\alpha_k\}$. The arrival process
$\{\tau_k\}$  has an intensity function $\lambda(\tau)$ referred to as
the (path) arrival rate \cite{turin}.
For the  most often considered case of uncorrelated
zero mean gains, the second moment of the received signal reads 
\begin{equation}
  \label{eq:93}
  \mathds E [|y(\tau)|^2 ] = \int_{-\infty}^{\infty} P(\tau-t) |s(t)|^2 dt, 
\end{equation}
where the power-delay  spectrum, $P(\tau)  $, is a product 
\begin{equation}
  \label{eq:67}
P(\tau)  = \sigma_\alpha^2(\tau)
\lambda(\tau), 
\end{equation} 
where $\sigma_\alpha^2(\tau)$  denotes the variance  of a complex gain
at a given the delay. A particularly
prominent example is the model by Turin  \cite{turin} where the delays
are drawn from a Poisson point process.  Although Turin's model was
originally intended for urban radio channels, it has since
been taken as basis for a wide range of models for outdoor and indoor channels including the models by, Suzuki~\cite{suzuki},  Hashemi~\cite{hashemi},
Saleh and Valenzuela \cite{saleh}, Spencer et al. \cite{Spencer2000} and
Zwick et al. \cite{Zwick2002,Zwick2000}. More recently, this type of statistical
channel models  has been considered for the millimeter-wave spectrum \cite{Haneda2015,Samimi2016}. 


Simulations and analyses based on a model are only trustworthy if the
parameter settings are properly chosen. For this, 
empirical methods for estimation of parameters are
wide-spread in the literature. Indeed, Turin along with the
scientists elaborating this modeling approach 
\cite{suzuki,hashemi,saleh,Spencer2000,Zwick2002,Zwick2000,Haneda2015,Samimi2016}
determined the parameters based on measurements. The empirical
approach, however, gives only limited insight into how model parameters
are affected by the propagation environment or system
parameters such as frequency bands and antenna
configurations. Therefore, costly measurement campaigns performed
to determine model parameters for one type of environment for one type
of radio system may have to be redone in case the model should be
adapted to a different situation, e.g. if considering new frequency bands or
different antenna configurations.  As a much less explored
alternative to the empirical approach, model parameters can in some
cases be obtained by analysis of the 
propagation environment. Potentially, this analytical approach allows
us to predict how changes in the propagation environment or in system
parameters will affect the channel model parameters.
Unfortunately, realistic propagation environments are
often too complex to permit such analysis and therefore, we can at best hope to
analyze simplistic, but elemental, scenarios. Such elemental
results may help us to better understand more realistic scenarios.

The elemental case where one transmitter and one receiver are situated
in the same rectangular room has been studied in a
number of works
\cite{Holloway1999,Rudd2003,Rudd2007,Andersen2007,Bamba2012,Steinbock2015}
channel using the theory of room electromagnetics inspired by the well
developed theory of room acoustics \cite{Kuttruff2000}.  This scenario
is relevant since many rooms in old and modern buildings are indeed
rectangular.  These investigations have focused on determining the
reverberation time which characterizes the exponentially decaying
reverberation tail of the average power-delay profile, or power-delay
spectrum. Room electromagnetics has also been considered as a means to
set the parameters of other models through entities derived from the
power-delay spectrum \cite{Steinboeck2013d,Steinboeck2016,Steinboeck2013}.


Models of the type \eqref{eq:66} are unidentifiable in the power-delay
spectrum---according to \eqref{eq:67} exactly the same delay-power spectrum can be obtained by a continuum of
combinations of arrival rates and conditional mark variances. This
effect is clearly present for Turin's model, but as noted in
\cite{Jakobsen2012}, also holds true for the Saleh-Valenzuela model \cite{saleh}:
by interchanging inter- and intra-cluster parameters for rates and
complex gains, and thereby completely altering the model's behaviour, the same
power-delay spectrum is obtained.  If two
of the three entities related through \eqref{eq:67}, are
specified, the third can be determined. While validated
room-electromagnetic models for the delay-power spectrum are already
available in the
literature. e.g. \cite{Holloway1999,Rudd2003,Rudd2007,Andersen2007,Bamba2012,Steinbock2015},
the  room electromagnetic modeling of the arrival rate appears to be still unexplored.

Here, we propose to model for the arrival rate by  an 
analysis  inspired by Eyring's  model \cite{Eyring1930} for reverberation
time in room acoustics. Eyring's model is developed for prediction of 
reverberation time in rooms with large average
absorption coefficient which is the typical situation situation in
room electromagnetics  \cite{Steinbock2015}.  Interestingly, in the
process of deriving the reverberation time using an approximation based on mirror
source theory, Eyring actually derived an approximation for
the arrival rate at large delays for a rectangular room using mirror
source theory. According to Eyring's
approximation the arrival rate increases quadratically with delay and
is inversely proportional to the room volume. This model thus captures a  transition effect
of  the received signal from early specular contributions to the late
diffuse reverberation tail, similar to the effect considered for
in-room radio propagation
\cite{Kunisch2003a,Kunisch2002,Pedersen2012,Pedersen2007}. 
Eyring's model for the reverberation
time has been recently considered and experimentally validated within
room electromagnetics\cite{Holloway1999,Steinbock2015} but his results on the
arrival rate has not yet been noticed or utilized.   

The contributions of the
present paper is to adapt Eyring's analysis to radio channel modeling by
including random antenna positions and orientations, as well as to
account for antenna directivity.  The effect of the antenna
directivity on the ``richness'' of measured impulse responses has
been noticed qualitatively  in early measurements \cite{Manabe1996} and  the impact of
antenna directivity on small scale fading parameters has beed studied
in several works \cite{Goodman2006,Yang2008,Smulders2009}. 
Our approach leads to an exact expression for the mean arrival rate
for the mirror source model;  for special cases our expression coincide with
Eyring's approximation. The rate is quadratic in delay, inversely
proportional to the room volume, and proportional to the product of
beam coverage fractions of the transmitter and receiver antennas. Making use of the exact
formula, we characterize the onset of the
diffuse tail by defining a ``mixing time'' as the point in time
where the arrival rate exceeds one component per transmit pulse
duration.  We also give an approximation for the power-delay
spectrum and study the mean delay and rms delay spread via 
simulations. It turns out that the
power-delay spectrum is   unaffected by the antenna
directivity, while the mean delay and rms delay spread vary.

We proceed in Section~\ref{sec:cons-rect-room} by introducing the
rectangular room considering non-isotropic
antennas for which we in Section~\ref{sec:mirrorSourceTheory} detail
the mirror source theory. Based on this model, we give approximations in
Section~\ref{sec:ArrivalRateAnalysis} for the arrival rate and
power-delay spectrum. In Section~\ref{sec:expect-arriv-count} we analyze the
mean arrival rate and power-delay spectrum for random transmitter position and
antenna orientation.  In Section~\ref{sec:simul-stud-exampl} we
illustrate the results of the analysis by Monte Carlo
simulations. Section \ref{sec:conclusion} concludes upon our contributions. 



\section{Considered Rectangular Room and Antennas}
\label{sec:cons-rect-room}
\begin{figure}
  \centering
\includegraphics[width=0.8\linewidth]{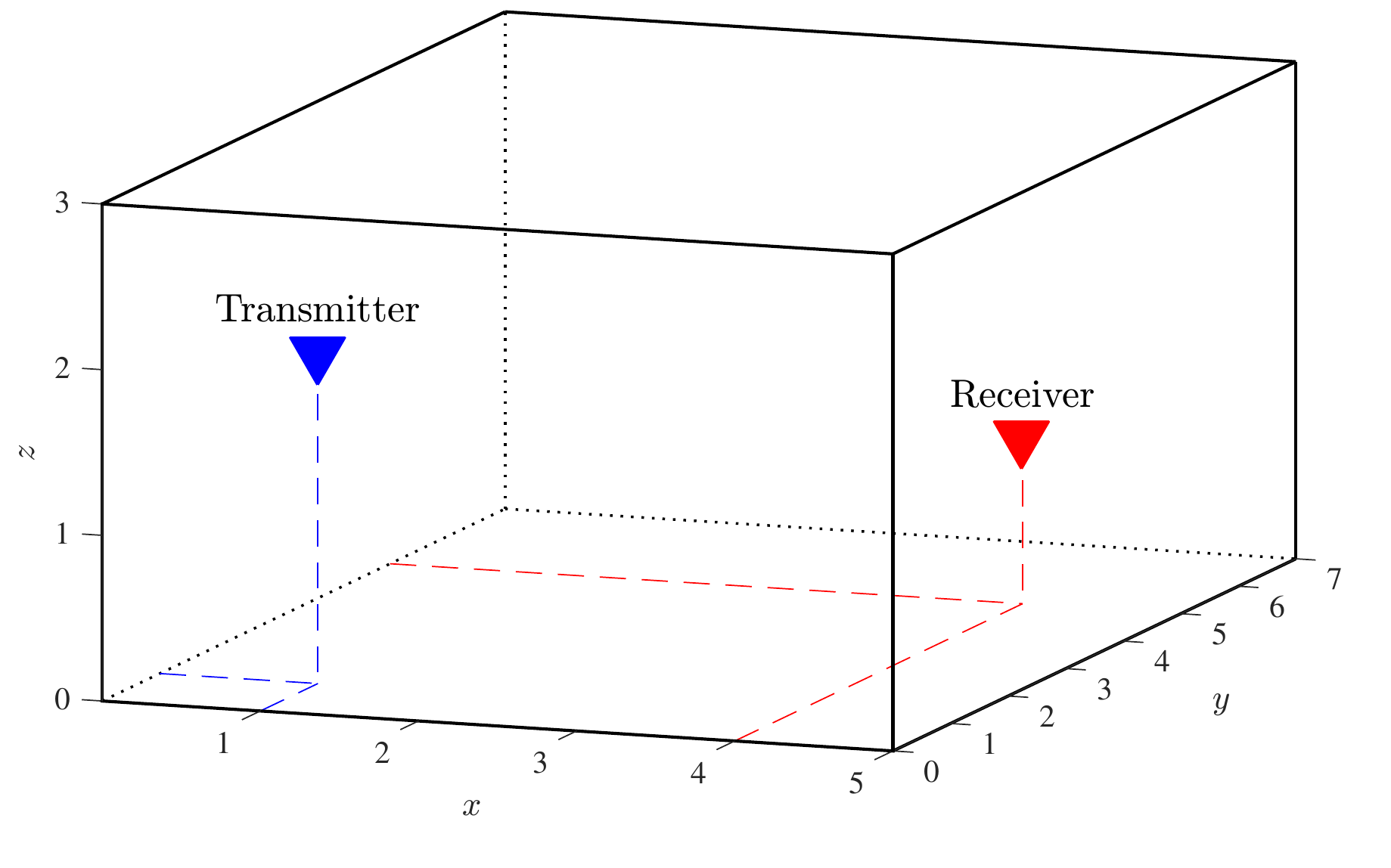}
  \caption{Three-dimensional rectangular room with transmitter and
    receiver inside along with the coordinate system.}
 \label{fig:room}
\end{figure}

Consider a rectangular room illustrated in Fig.~\ref{fig:room} with
directional  transmitter and receiver antennas located inside. 
The room is of dimension $L_x\times
L_y \times L_z$ and has volume $V=L_xL_yL_z$. The six walls of the
room (including floor and ceiling) denoted by
$W_1,\dots,W_6$.  We assume  that the carrier wavelength $l_c$ is small
compared to the room dimensions, and that only specular reflections
occur with a gain independent of incidence direction.  The power gain
(or reflectance) of wall $i$ is denoted by $g_i$ to. Positions are given with reference to a
Cartesian coordinate system with origin at one corner and aligned such
that the room spans the set $[0,L_x)\times[0,L_y)\times[0,L_z)$.  Then
the positions of the transmitter and receiver are given as $
r_T = [x_T,y_T,z_T]^T$ and $r_R = [x_R,y_R,z_R]^T$. 

To describe the directive antennas, we now introduce a simplistic
model. We only define the notations for the
transmitter antenna, indicated by subscript $T$; similar entities
for the receiver antenna have subscript $R$. For simplicity reason, we
ignore polarization of the antennas and describe these by only a
directional power gain. We denote the antenna
gain by $G_T(\Omega)$ (power per solid angle) where the direction
specified by the unit vector $ \Omega \in \mathbb S_2$. To simplify
notation, we assume the antennas to be lossless, and thus the integral
of the antenna gain over the sphere equals $4\pi$. \footnote{The  forthcoming
analysis does not change substantially by considering lossy
antennas. The equations can be readily adapted to by including  the radiation efficiency
in equations where the antenna gain enters.}  The beam support  is the
portion of the sphere, denoted by $\Omega_T$, is defined as the support of the
function\footnote{Alternatively, one may define the beam coverage
  solid angle as the portion of the sphere where $G_T(\Omega)$ exceeds
  a specified value. } $G_T(\Omega)$:
\begin{equation}
  \label{eq:70}
 \Omega_T = \{ \Omega: G_T(\Omega)\neq 0\}, 
\end{equation} With this definition, the beam coverage solid angle of
the transmitter antenna ranging from  zero to $4\pi$ reads
\begin{equation}
  \label{eq:26}
|\Omega_T| =   \int_{\mathbb S_2} \mathds 1 (\Omega \in \Omega_T )  d\Omega,
\end{equation}
where $\mathds 1(\cdot)$ denotes an indicator function with 
value one if the argument is true and zero otherwise. 
To shorten the notation, we further define the beam coverage fraction as
 \begin{equation}
   \label{eq:90}
   \omega_T = \frac{|\Omega_T|}{4\pi}
 \end{equation}
 The beam coverage fraction ranges from zero to one and can be
 interpreted as the probability that a wave impinging from a uniformly
 random direction is within the antenna beam.


\section{Mirror Sources and Multipath Parameters}
\label{sec:mirrorSourceTheory}
For the defined setup, mirror source theory predicts that the received signal is an infinite sum of
attenuated, phase-shifted and delayed signal components as in
\eqref{eq:66}. Unlike Turin's model, in this case the pairs of
delay and complex amplitudes $\{(\tau_k,\alpha_k)\}$ do not form a
marked Poisson process but are given by the geometry of the
propagation environment. The complex gains and delays are readily described using the
theory of mirror sources as follows.



\begin{figure}










\centering 
\includegraphics[width=0.8\linewidth]{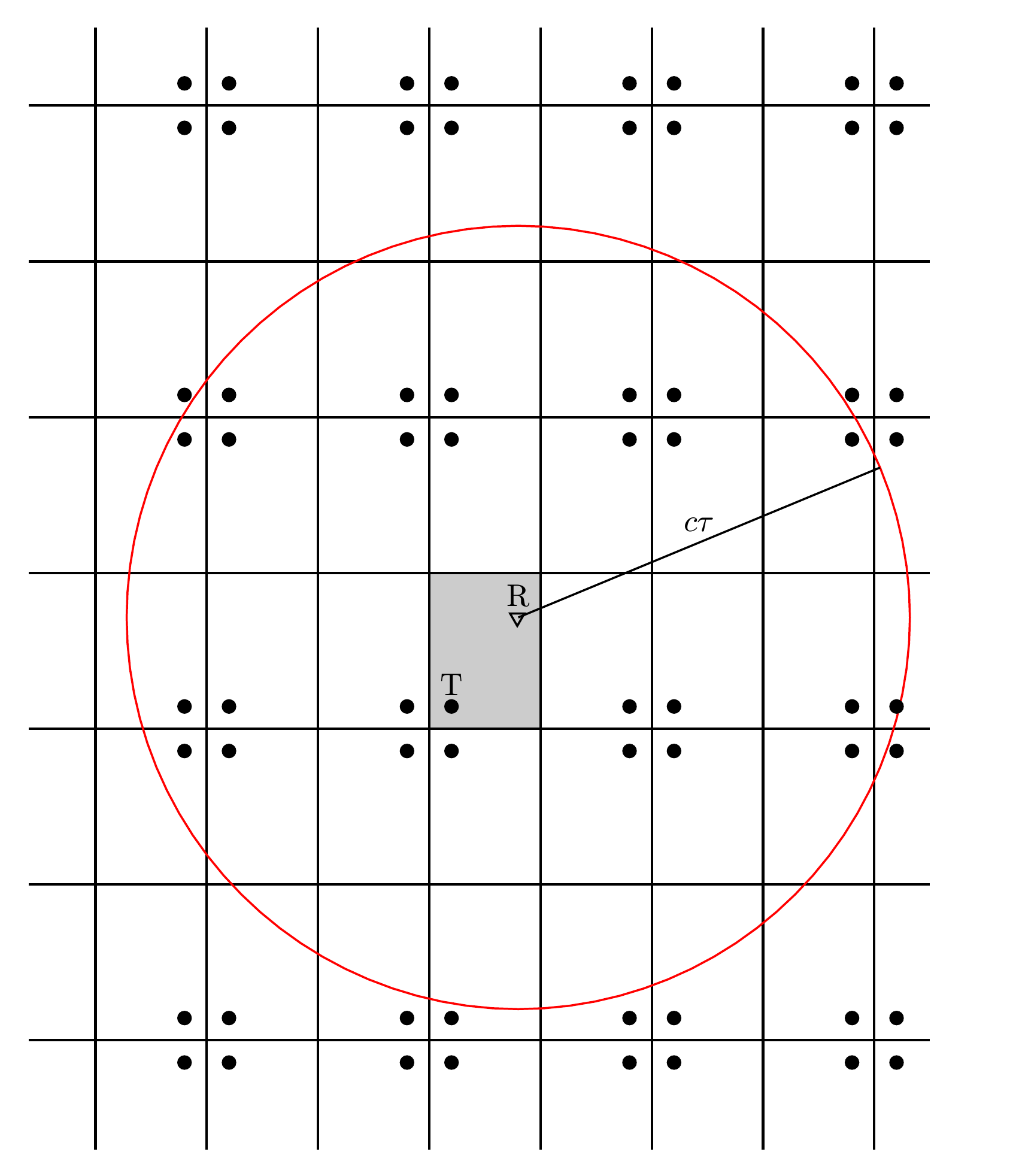}
  \caption{The rectangular room seen from above  with 
    transmitter $T$ and a receiver $R$  and a number of mirror
    rooms. The pattern continues similarly in the direction
    perpendicular to the drawing plane. Contributions from mirror
    sources inside the sphere of radius $c\tau$ arrive at the receiver at delays smaller than $\tau$.}
  \label{fig:mirrorRoom}
\end{figure}

To construct the path from $T$ to $R$ via a single reflection at wall $W$ we
determine the position of mirror source $T'$ by
mirroring $T$ in $W$.  Thereby, the interaction point can be
determined as the wall's intersection with the straight line segment
from $T'$ to $R$.
The two-bounce path $T-W_1-W_2-R$ may be constructed by mirroring
 wall $W_1$ in wall $W_2$ to construct $W_1'$ and then mirroring
$T'$ in $W_1'$. By repeating this procedure \emph{ad infinitum} we can
construct and  infinite set of mirror  sources and mirror rooms 
as illustrated in Figure~\ref{fig:mirrorRoom}. The position of mirror
source $k$ can be computed as
\begin{equation}
  \label{eq:68}
r_{T(k_x,k_y,k_z)}
 = 
\begin{bmatrix}
\big\lceil \tfrac{k_x}{2} \big\rceil \cdot 2 L_x + (-1)^{k_x}\cdot x_{T}\\
\big\lceil \tfrac{k_y}{2} \big\rceil \cdot 2 L_y + (-1)^{k_y}\cdot y_{T}\\
\big\lceil \tfrac{k_z}{2} \big\rceil \cdot 2 L_z + (-1)^{k_z}\cdot z_{T}  
\end{bmatrix}
\end{equation}
where $k_x$ is the number of reflections on the two walls parallel to
the $y-z$-plane, and similarly for $k_y,k_z$. Hence, the path index path
index $k$ corresponds to a triplet $k=(k_x,k_y,k_z)$.
Alternatively, the same path can be constructed by introducing a
\emph{mirror receiver} at position $r_{Rk}$ determined by 
replacing subscript $T$ by subscript $R$ in \eqref{eq:68}.
Notice the direct (or line-of-sight) path is also included for $k=(0,0,0)$, since
for this case $r_{T(0,0,0)} = r_T$ and $
r_{R(0,0,0)} = r_R$. For notational brevity, we use subscript $k=0$
instead of $k=(0,0,0)$  for entities related to the direct path
throughout the paper.

The signal emitted by mirror source $k$  arrives at the receiver with  delay
$\tau_k$. Analogously, the signal emitted by the transmitter 
observed by mirror receiver $k$ has the same delay $\tau_k$.
The delay of path $k$ be computed from the positions of
mirror source $k$ or mirror receiver $k$ as
\begin{equation}
  \label{eq:69}
\tau_k = \|  r_{Tk} - 
r_R\|/c = \|r_{Rk} - r_T\|/c,  
\end{equation}
where $c$ is the speed of light. 

The directions of departure and arrival for each path can also be
computed.  The direction of arrival of the signal from mirror source
$k$ is given by the unit vector
\begin{equation}
  \label{eq:41}
 \Omega_{Rk} = \frac{ r_{Tk}- r_R}{\| r_{Tk}- r_R\|}.
\end{equation}
Similarly, the direction of departure of path $k$ denoted by
$\Omega_{Tk}$ and can be computed from \eqref{eq:41} by interchanging
subscripts $T$ and $R$.  It follows that
directions of departure and arrival of a specific path are related as
\begin{align}
  \Omega_{Tk}&=
-
\begin{bmatrix}
    (-1)^{k_y+k_z} && \\
    &(-1)^{k_x+k_z} & \\
    &&(-1)^{k_x+k_y}  
  \end{bmatrix}
\Omega_{Rk}.
\end{align}
In particular, for the direct path  $\Omega_{T0} =
-\Omega_{R0}$.

Finally, the  gain of path $k$ can be specified. We shall not be
concerned with the phase of the complex gain $\alpha_k$, but only its
squared magnitude, i.e. the corresponding power gain.  The power
gain of path $k$ reads 
\begin{align}
  \label{eq:147}
  |\alpha_{k}|^2 &= g_k \cdot \frac{G_T(\Omega_{Tk})
    G_R(\Omega_{Rk})}{(4\pi c\tau_k / l_c)^2} 
\end{align}
where the factor $g_k$ denotes the gain due to reflections on the
walls, the numerator accounts for the transmitter and receiver
antennas, and  the denominator is due to the attenuation of a spherical wave
with $l_c$ denoting the wavelength.
Naming the walls parallel to the $yz$-plane as $W_1$ and $W_2$
respectively, we see that path $k$ interacts with wall $W_1$ in total
$\big|\lfloor\tfrac{k_x}{2} \rfloor\big|$ times and with wall $W_2$ in
total $\big|\lceil\tfrac{k_x}{2} \rceil\big|$ times. The numbers of
interactions with  other walls are computed similarly. Consequently,
\begin{multline}
  \label{eq:148}
g_k =
  g_1^{\big|\lfloor\tfrac{k_x}{2} \rfloor\big|}
  g_2^{\big|\lceil\tfrac{k_x}{2} \rceil\big|}
  g_3^{\big|\lfloor\tfrac{k_y}{2} \rfloor\big|}
  g_4^{\big|\lceil\tfrac{k_y}{2} \rceil\big|}
  g_5^{\big|\lfloor\tfrac{k_z}{2} \rfloor\big|}
  g_6^{\big|\lceil\tfrac{k_z}{2} \rceil\big|}.  
\end{multline}
Henceforth, we consider for simplicity all walls to have the
same gain value $g=g_1=\dots = g_6$. Then the gain of path $k$ 
simplifies as $g_k = g^{|k|} $ with the convention $|k| = |k_x|+|k_y|+|k_z|$.
Furthermore, we remark that for the direct path the
expression \eqref{eq:147} reduces to the Friis equation \cite{Friis1946} for
propagation in free space.

\subsection{Numerical Examples}
\label{pdpExamples}
Before proceeding with analysis of the mirror source model, we
first illustrate how the received signal behaves with a few numerical
examples. The settings are specified 
in Table~\ref{tab:simulationSettings}. The transmitter and receiver  have identical sector
antennas. The transmit antenna gain is
\begin{equation}
  \label{eq:86}
  G_T(\Omega) = \frac{1}{\omega_T}
\mathds 1(\Omega^T\zeta_T   \geq 1-2\omega_T),
\end{equation}
i.e. the gain is constant over the spherical cap centered at the
direction given by the unit vector $\zeta_T$. This implies a half-beam
width of $\arccos (1-2\omega_T)$.  The receiver antenna is defined
similarly.  In this example, we orient the antennas in direction of
line-of sight, i.e. $\zeta_T = \Omega_{T0}$ and
$\zeta_R = \Omega_{R0}$.  The transmitted signal $s(t)$ is a sinc
pulse which has constant Fourier transform over the considered
frequency bandwidth, and zero elsewhere.
\begin{figure}
\centering 
\includegraphics[width=\linewidth]{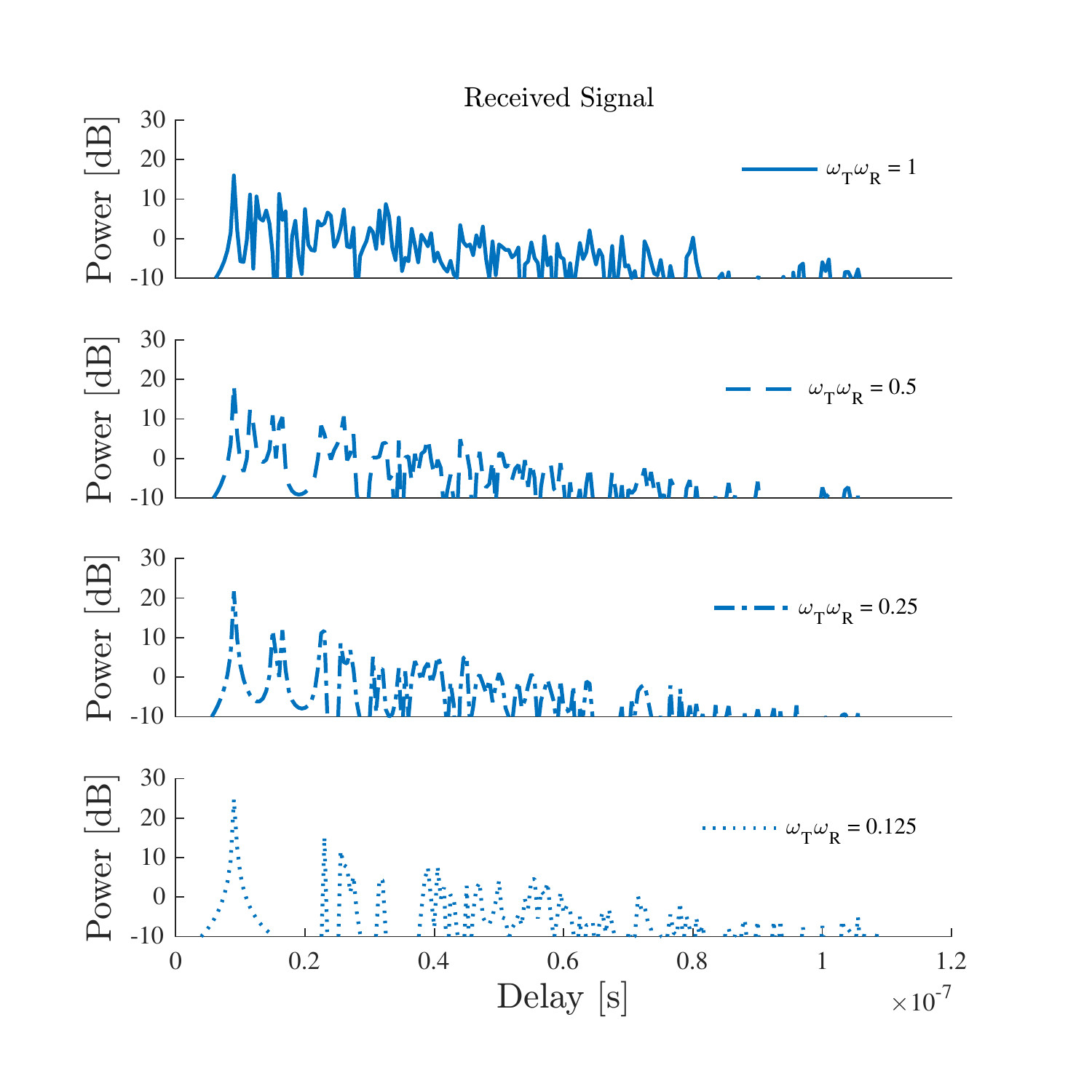} 
\caption{Examples of magnitude square of received signal for four different
   $\omega_T\omega_R$ values for the setup  given in Table~\ref{tab:simulationSettings}. The antennas are located at $ r_T = [2.5,2.5,1.5]^T\meter$ and $
   r_R = [3.8,4.0,0.6]^T\meter$ and are oriented exactly in the
   direction of line-of-sight. }
  \label{fig:receivedSignal}
\end{figure}

\begin{table}
  \centering
  \caption{Simulation Settings }
  \begin{tabular}{lc}
\toprule
  Room dim., $L_x\times L_y\times L_z $& $5\times 5\times 3\, \cubic\meter$\\
  Reflection gain, $g$ & $0.6$\\
 Center Frequency & $60\,\giga\hertz$\\
 Bandwidth, $B$& $2\,\giga\hertz$\\
 Speed of light, $c$ &$3\cdot 10^8\,\meter\per\second$\\ 
\bottomrule 
\end{tabular}
  \label{tab:simulationSettings}
\end{table}

Fig.~\ref{fig:receivedSignal} shows received signals for four different antenna settings; isotropic
antennas ($\omega_T\omega_R = 1$), and three cases with directive
antennas. The general trend is that the received signal decays exponentially with
delay while the signal contributions gradually merge into a diffuse
tail. The rate of diffusion depends on the antenna directivity: higher
antenna directivity leads to a slower diffusion process.

\section{Analysis of Deterministic Mirror Source Model}
\label{sec:ArrivalRateAnalysis} 
The equations \eqref{eq:68}--\eqref{eq:148} define
the mirror source model to an extent where the model can be simulated
from, but are difficult to interpret directly. To better understand the behavior of
the model we next consider  approximations for the arrival count, arrival
rate and power-delay spectrum. In this section assume the
antenna positions and orientations as deterministic. Later, in
Section~\ref{sec:expect-arriv-count}, we randomize these variables.
\subsection{Arrival Count} 
\begin{figure}
\centering 
\includegraphics[width=\linewidth]{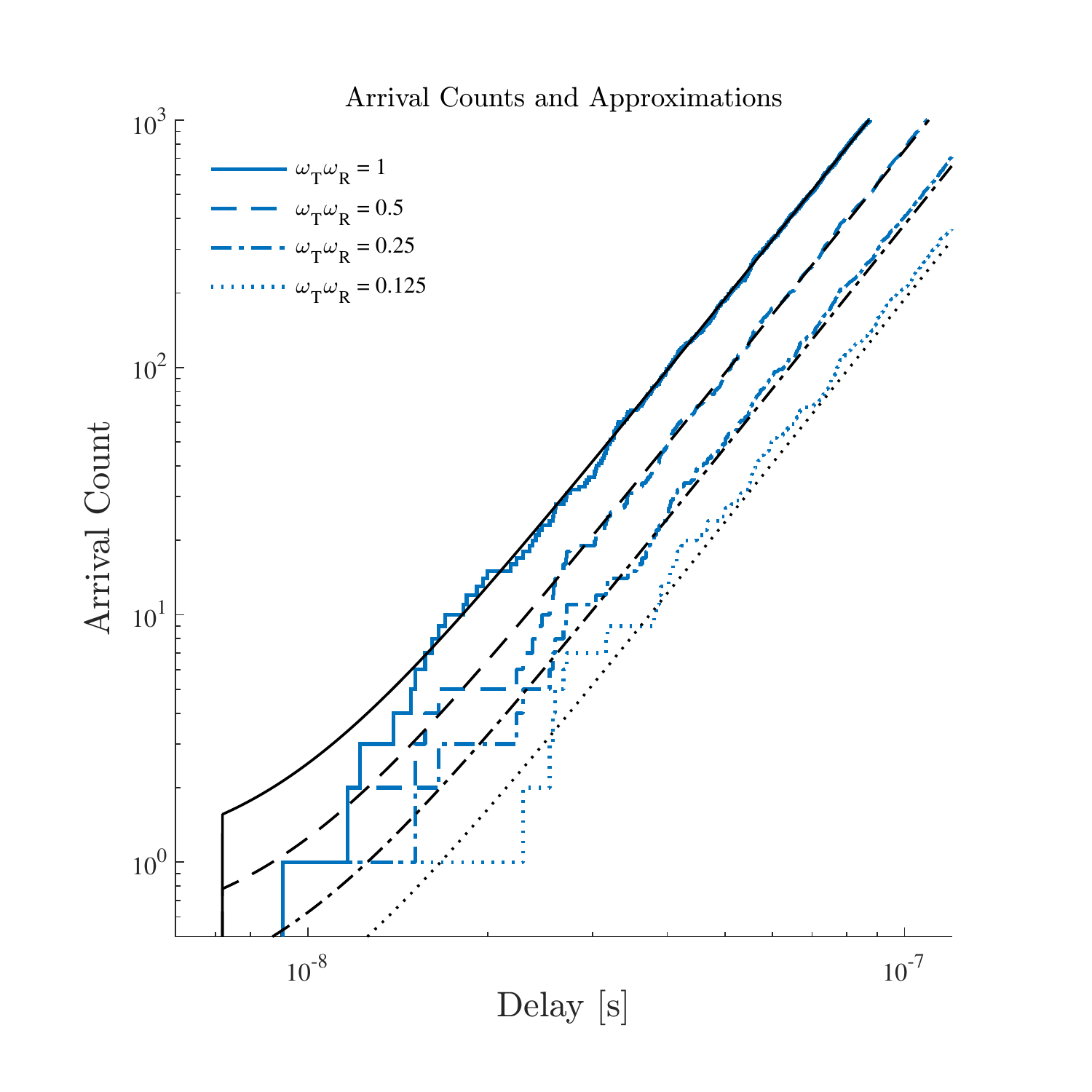} 
\caption{Arrival counts (blue lines) and corresponding approximation 
   obtained from \eqref{eq:53} (black lines) corresponding to the
   examples in Fig.~\ref{fig:receivedSignal}. }
  \label{fig:arrivalProcess}
\end{figure}

The arrival count $N(\tau)$ is defined as the number of paths contributing to the received
signal components up to and including a certain time $\tau$. For a path to
contribute, the corresponding mirror source  should be within a
radius $c\tau$ of the receiver, see
Fig.~\ref{fig:mirrorRoom}.  Furthermore, the path should be within the
beam support of both the transmitter and receiver antennas. Thus the 
arrival count can be expressed as
\begin{align}
  \label{eq:54}
  N(\tau)
&= \sum_{k} \mathds  1(\tau_k\leq\tau  ) \cdot 
\mathds 1(\Omega_{Tk}\in\Omega_T) 
 \cdot \mathds 1( \Omega_{Rk}\in\Omega_R ) 
\end{align}
The count depends on the antenna positions, orientations and as
exemplified in Fig.~\ref{fig:arrivalProcess}, on the
specific antennas. As can be seen from the example, the
 count  approaches a cubic asymptote for large delays. 
This cubic trend was  noticed and approximated by Eyring as
\cite{Eyring1930}
\begin{align}
  \label{eq:57}
N(\tau) \approx \frac{4\pi c^3\tau^3 }{3V}, \qquad \tau\gg0.
\end{align}



We now develop an approximation for the arrival count by adapting
Eyring's analysis to the current setting. First, we approximate the term due to the direct
path as\footnote{If this approximation is not made, a more accurate
  expression is obtained at the expense of introducing dependency of
  antenna orientations in the overall a approximation of the arrival count which
  is undesirable here.} 
\begin{equation}
  \label{eq:98}
\mathds  1(\tau_0\leq\tau  ) \omega_T \omega_R
\end{equation}
Secondly,  the number of mirror sources with delay less than $\tau$ equals
the number of mirror sources inside the sphere with radius
$c\tau$ centered at the receiver. For $c\tau$  large compared to 
the diagonal of the room, i.e. $ c \tau \gg
\sqrt{L_x^2+L_y^2+L_z^2} $,  the number of such mirror sources is approximately  
\begin{equation}
  \label{eq:99}
\frac{4\pi c^3(\tau^3-\tau_0^3)}{3V}  
\end{equation}
where we include one mirror source per room volume and subtract the
contribution due to the volume closer than $c\tau_0$ to the receiver. Fourthly, only a fraction,
$\omega_R$, of these mirror sources are picked up by the
receiver. Ignoring the dependency between the direction of
departure and arrival for indirect components, we account for the
transmit antenna by a factor $\omega_T$. From this reasoning we have
\begin{align}
  N(\tau) & \approx 
\mathds  1(\tau\geq \tau_0 ) \left[1+  \frac{4 \pi c^3(\tau^3-\tau_0^3)}{ 3 V}\right]
\omega_T  \omega_R.
  \label{eq:53}
\end{align}
As exemplified in Fig.~\ref{fig:arrivalProcess}  this approximation follows closely 
the asymptote of the exact count. For the special case of isotropic and colocated  antennas
expression  \eqref{eq:53}  equals  Eyring's approximation \eqref{eq:57} plus one.

\subsection{Arrival Rate}
The arrival rate, denoted by $\lambda(\tau)$, is expected  number of
signal components arriving at the receiver per unit time at delay
$\tau$  which can be defined it terms of the arrival count such that the
expression
\begin{align}
  \label{eq:58}
 \mathbb E[ N(\tau)] = \int_{-\infty}^{\tau} \lambda(t) dt
\end{align}
holds true. Essentially, $\lambda(\tau)$ can be thought of as a
``derivative'' of $\mathbb E [N(\tau)]$. However, since the count $N(t)$
is deterministic, we see that  $\mathbb E [N(\tau)]$ equals 
$N(\tau)$. The count is a step function and therefore  $\lambda(\tau)$
should be interpreted in the distribution sense as a Radon-Nikodym
derivative (with respect to Lebesque measure) which leads to: 
 \begin{equation}
   \lambda(\tau) = 
 \sum_{k} \delta (\tau-\tau_k  ) \cdot 
 \mathds 1(\Omega_{Tk}\in\Omega_T) 
 \cdot \mathds 1( \Omega_{Rk}\in\Omega_R ), 
 \end{equation}
where $\delta(\cdot)$ is Dirac's delta function.
Again, the exact count yields no valuable interpretation. 
Instead by inserting the approximation
\eqref{eq:53} for the arrival count into \eqref{eq:58} we approximate the arrival rate
as 
\begin{align}
\lambda(\tau) \approx 
\delta(\tau-\tau_0) \omega_T  \omega_R   +\mathds 1 (\tau>\tau_0) 
 \frac{4\pi c^3\tau^2 }{ V} \omega_T  \omega_R .  
  \label{eq:59}
\end{align}
The approximation
\eqref{eq:59} is clearly not valid point-wise, but should be seen as the average
number of arrivals per time unit in a small time interval centered at $t$.

The expression \eqref{eq:59} gives rise to a number of observations.
First, the arrival rate is quadratic in delay which is in sharp
contrast to the widespread Saleh-Valenzuela model \cite{saleh} where
the delays of each ``cluster'' of components has constant arrival
rate. Moreover, considering that clusters also arrive at constant
rate, the overall arrival rate is only linearly increasing in delay
\cite{Jakobsen2012}.  Secondly, the arrival rate in \eqref{eq:59} is
inversely proportional to the room volume for $\tau> \tau_0$. Thus,
larger rooms lead to smaller arrival rates. This implies that attempts
to empirically characterize arrival rates for inroom channels should
pay attention to the room size. Finally, we observe that the antennas
affect the arrival rate by a delay-independent scaling. Thus very
directive antennas lead to a sparser channel in the early part of the
channel response, in agreement with experimental results presented in
\cite{Manabe1996,Goodman2006,Yang2008,Smulders2009}.  However, the
arrival rate still grows quadratically and eventually the components
in the response merge into a diffuse tail.


\subsection{Approximation for Power-Delay Spectrum }
We now derive an approximation of the power-delay spectrum. Eyring
noted in \cite{Eyring1930} that the number of wall
interactions for mirror source $k$ is roughly 
\begin{equation}
  \label{eq:77}
|k| \approx \tau_k \cdot \frac{cS }{4V}, \quad k\neq (0,0,0).
\end{equation}
with $S=2(L_xL_y + L_xL_z+L_yL_z)$ denoting the surface area of the
room. Inserting this into~\eqref{eq:147}  yields (for $g=g_1=\dots = g_6$)
\begin{equation}
  \label{eq:78}
    |\alpha_k|^2\approx
    \begin{cases}
 g^{ \tau_k  cS/4V} \cdot \frac{G_T(\Omega_{Tk}) 
    G_R(\Omega_{Rk})}{(4\pi c\tau_k/l_c)^2},& k\neq (0,0,0) \\
 \frac{G_T(\Omega_{T0}) 
    G_R(\Omega_{R0})}{(4\pi c\tau_0/l_c)^2},& k= (0,0,0)       
    \end{cases}
\end{equation}
For propagation paths with large delay, the direction of departure and
arrival are close to uniformly distributed on the sphere. Thus we
approximate the gain due to the transmitter antenna for a direction of departure within the beam coverage solid angle as
\begin{equation}
  \label{eq:28}
 \frac{1}{|\Omega_T|}\int_{\Omega_T} G_T(\Omega)   d\Omega =
 \frac{1}{|\Omega_T|}\int_{\mathbb S_2} G_T(\Omega)   d\Omega =\frac{1}{\omega_T}
\end{equation}
A similar expression is obtained at the receiver side.
Now, further assuming independent directions of departure and arrival, we have for the conditional second moment, 
\begin{equation}
  \label{eq:79}
  \sigma_\alpha^2(\tau ) \approx 
  \begin{cases}
\frac{g^{ \tau  cS/4V}}{(4\pi c\tau/l_c)^2}\cdot \frac{1}{\omega_T\omega_R}  
& \tau>\tau_0 \\
\frac{1}{(4\pi c\tau/l_c)^2}\cdot  \frac{1}{\omega_T\omega_R}  & \tau=\tau_0.     
  \end{cases}
\end{equation}
 Combining with the expression for the arrival rate, the approximation
for the power-delay spectrum reads
 \begin{align}
  \label{eq:74}
  P(\tau)\approx 
\delta(\tau-\tau_0) 
\frac{1 }{(4\pi c\tau_0/l_c)^2} 
 + \mathds 1 (\tau>\tau_0)  \frac{e^{-\tau/T} }{4\pi V/l_c^2 c}  
\end{align}
with the reverberation time $T$  defined as
\begin{equation}
  \label{eq:103}
  T =  -\frac{4V}{cS\ln(g)}.
\end{equation}

This expression for the power-delay spectrum is remarkable in a number
of ways.  First, the form of the power-delay spectrum appears to
be a spike plus an exponential tail. This is interesting in the light
of  the super-exponential decay of the per-path gain in \eqref{eq:79}.  
However, this super-exponential trend is balanced out by the quadratic
increase in arrival rate such that the net result is an exponential decay.  Second, the positions of the transmitter and receiver antennas only enter via the
delay of the direct component. This implies the expected result that
the decay rate of the tail is constant throughout  the whole room as
is well known in  room electromagnetics. However, the
onset of the tail depends on the delay of the direct component. This
exact structure was the one studied in great
detail in  \cite{Steinboeck2013}.  Third, the power-delay spectrum appears to be unaffected by the directivity of
the antenna. Indeed, the antennas enter in both the arrival rate and
in the conditional gain, but the effects cancel in the power-delay
spectrum.

The approximation in \eqref{eq:78} can be made more accurate by
incorporating more complex models such as the ones developed for room
acoustics, see \cite{Neubauer2001, Kuttruff2000}. As an example, the
modification introduced in \cite{Kuttruff2000} accounts for 
this discrepancy due to the approximation in \eqref{eq:77} where
a random variable is replace by its mean. This modification  amounts to adjusting the 
reverberation time by a correction factor $\xi$ defined as
\begin{equation}
  \label{eq:1}
\xi = \frac{1}{1+\gamma^2\ln(g)/2}. 
\end{equation}
where the constant $\gamma^2$,  which depends on the aspect ratio of the
room, can be determined by
Monte Carlo simulation and typically takes
values in the range 0.3 to 0.4 \cite{Kuttruff2000}.
The particularities of
such corrections are of less importance for the forthcoming
analysis and therefore further refinements of \eqref{eq:1}  are left
as future work.




\section{Analysis of Randomized Mirror Source Model}
\label{sec:expect-arriv-count}
In the previous section,  antenna positions and orientations were
held fixed. In the sequel, we let let the position and
orientation of the transmitter be random. 




\subsection{Mean Arrival Count and Arrival Rate}
Suppose that the position and orientation of the receiver antenna is
fixed. In contrast hereto, the transmitter's position is random with a
uniform distribution on the room, i.e. that $r_{T} \sim \mathcal U (
[0,L_x]\times[0,L_y]\times[0,L_z])$. Furthermore, let the
transmitter's orientation be random according to a uniform
distribution on the sphere. The counting variable $N(\tau)$ is random with
mean
\begin{align}
  \mathbb E[N(\tau)] &=\mathbb E\big[ 
 \sum_{k}  
\mathds  1(\tau_k<\tau  ) \cdot 
\mathds 1( \Omega_{Tk}\in \Omega_T )  \cdot 
\mathds 1( \Omega_{Rk}\in \Omega_R ) ].
\end{align}
Since the orientation of the transmitter antenna is uniformly random,
the probability for any particular fixed direction, to reside in the
random beam support $\Omega_T$, equals the beam coverage fraction
$\omega_T$. Thus, we have the conditional mean 
\begin{equation}
  \label{eq:5}
\mathbb E[\Omega_{Tk}\in\Omega_{T}|\Omega_{Tk}]=\omega_T
\end{equation}
irrespective of the particular value of $\Omega_{Tk}$.
Each mirror sources is uniformly distributed within its mirror room,
and therefore mirror source positions 
constitue a homogeneous (but not Poissonian) random spatial point
process  with intensity $1/V$. Then, inserting \eqref{eq:5} and using
Campbell's theorem\cite{Stoyan1995}, we can  rewrite the expectation as an integral
over mirror source positions 
 \begin{align}
  \mathbb E[N(\tau)]
&=\frac{\omega_T}{V}\int \mathds  1\left(\tfrac{\| r- r_R\|}{c}<\tau
\right) 
 \mathds 1\left(\tfrac{r-r_R}{\| r- r_R\|}\in
\Omega_R\right) d  r\notag \\
&=\frac{4\pi c^3\tau^3}{3V} \omega_T \omega_R \mathds 1 (\tau>0).
\label{eq:75}
\end{align}
Taking the derivative of the expected arrival count, we obtain the corresponding arrival rate 
\begin{equation}
  \label{eq:63}
  \lambda(\tau) = \frac{4\pi c^3\tau^2}{V} \omega_T \omega_R \mathds 1 (\tau>0).
\end{equation}

It follows   by simple modifications of the above argument that the
same results hold true for a number of  different cases:
\begin{enumerate}
\item  Uniform receiver orientation  and transmitter fixed orientation
  and uniform location.
\item  Either of the antennas are isotropic and either of the antenna
  location are uniform.
\item Transmitter position and orientation are uniform and
  independent of the receiver position and orientation.
\item Transmitter position and antenna orientation are uniform
  conditioned on the receiver position and orientation.
\end{enumerate}
Obviously, Case 4) implies Cases 1) through 3). Moreover, by
symmetry, any of the above results hold true if the transmitter and
receiver swap roles.

This quadratically increasing rate bears witness of the
gradual transition in the impulse response that consists of 
separate specular components at early delays to a late diffuse tail
consisting of myriads of specular components. 
We remark that for isotropic antennas Eyring's approximation (see \eqref{eq:57}) is equal to
our expression for the mean count. In this sense,  Eyring's
approximation is not only valid asymptotically, but is exact in
the mean. The inclusion of the beam coverage fractions is a natural
extension to the non-isotropic case.


The relative ease by which we derived the mean arrival count 
\eqref{eq:75} may lead us to think that perhaps also higher moments
could be easily derived. However, it proves much more challenging to
derive its second moment---in fact we have not been able to establish
an exact expression. We give an approximation in
Appendix~\ref{sec:appr-raw-second}.  Similarly, it is difficult to get
exact formulas if less randomness is introduced in the model. To that
end,   Appendix~\ref{sec:mean-arrival-count} gives an upper bound to the
mean arrival rate for the case where the transmitter antenna has
random position, but fixed orientation; 
Appendix~\ref{sec:determ-transm-rece} gives an approximation for the
mean count for fixed transmitter-receiver distance.

\subsection{Mixing Time}
The arrival rate gives us a way to quantify the onset of this diffuse tail, i.e. the point in
time from which the preceding signal components merge together and can
no longer be distinguished. In analogy with room acoustics literature,
see \cite{Lindau2010}, we refer to this time as the ``mixing
time'' which we denote by $\tau_{\mathrm{mix}}$. 
For a system with signal bandwidth $B$ in which the receiver 
can distinguish on average one\footnote{The number  $N_{\mathrm{mix}}$ of components that can 
be distinguished within a pulse duration depends on the particular system in 
question. Determining this value is beyond the scope of the present 
investigation.  However, determining the mixing time in \eqref{eq:80}
for a $N_{\mathrm{mix}}$ different from unity results in a scaling by
$\sqrt{N_{\mathrm{mix}}}$.} signal component per pulse duration
$1/B$,  then we have $\lambda(\tau_{\mathrm{mix}})
= B$, and the mixing time can be expressed as
\begin{equation}
  \label{eq:80}
  \tau_{\mathrm{mix}}= 
\sqrt{\frac{BV}{4\pi c^3 \omega_T\omega_R}}.
\end{equation}
The mixing time is proportional to the square root of the room volume
which is quite intuitive: larger rooms have longer mixing
times. Moreover, the mixing time is inversely proportional to the
square root of the beam coverage fractions: more directive
antennas lead to a later onset for the diffuse tail. Finally,
by increasing the system bandwidth, the mixing time increases by the
square root of the bandwidth. The mixing time 
determines if a diffuse reverberation tail can be observed in noise
limited measurements. The diffuse tail appears only if the power-delay
spectrum exceeds the noise floor at the mixing time, and is otherwise
masked by noise.

\begin{figure}
\centering 
\includegraphics[width=\linewidth]{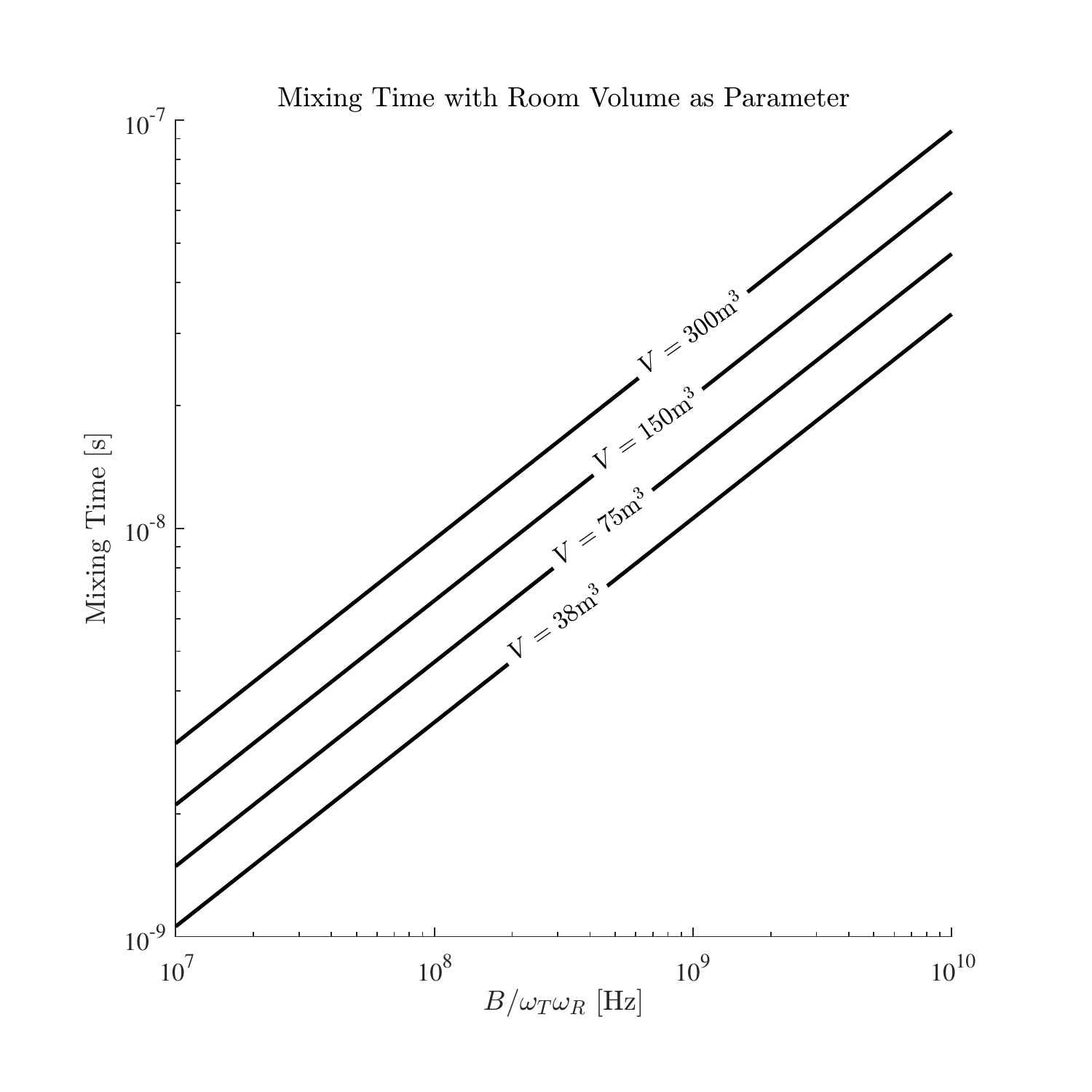}
 \caption{Mixing time versus $B/\omega_T\omega_R$ for different room volumes.}
  \label{fig:mixingTime}
\end{figure}

As an  example, the setup with the settings given in
Table~\ref{tab:simulationSettings} gives a mixing time of $21\
\nano\second$  for isotropic antennas and $42\ \nano\second$ for
hemisphere antennas. Fig.~\ref{fig:mixingTime} shows the mixing time versus
$B/\omega_T\omega_R$ for a range of room volumes.

\subsection{Approximation for Power-Delay Spectrum}
By following  essentially the same steps leading to the approximation
\eqref{eq:74} we can derive an approximation for the power-delay
spectrum. For simplicity, however, we ignore here the different gain of the direct
path and assign instead the same gain as any other path. Thus the
conditional second moment for the path gain reads
\begin{equation}
\label{eq:3}
 \sigma_\alpha^2(\tau ) \approx 
\frac{g^{ \tau  cS/4V}}{(4\pi c\tau/l_c)^2}\cdot \frac{1}{\omega_T\omega_R}  
\end{equation}
Combining this with the arrival rate in \eqref{eq:63}, we obtain
\begin{align}
\label{eq:2}
  P(\tau)\approx  \mathds 1 (\tau>0)  \frac{e^{-\tau/T} }{4\pi V/l_c^2 c} .
\end{align}
with the reverberation time $T$ defined in \eqref{eq:103} with
correction factor in \eqref{eq:1}.

\section{Simulation Study}
\label{sec:simul-stud-exampl}
We now illustrate  the theoretical results derived in the
previous sections by comparing them with Monte Carlo simulations.
We use nearly the same setup as in the numerical examples provided
earlier in Section~\ref{pdpExamples} with the same settings listed in
Table~\ref{tab:simulationSettings}. Compared to the previous setup,
there are two differences: 1) the transmitter  and receiver are
placed uniformly at random within the room and 2)  orientations picked
uniformly at random,  i.e. we simulate the setup leading to
 \eqref{eq:75} and \eqref{eq:63}. We perform 10\,000  Monte Carlo
 simulation runs.

\begin{figure}
  \centering
\includegraphics[width=\linewidth]{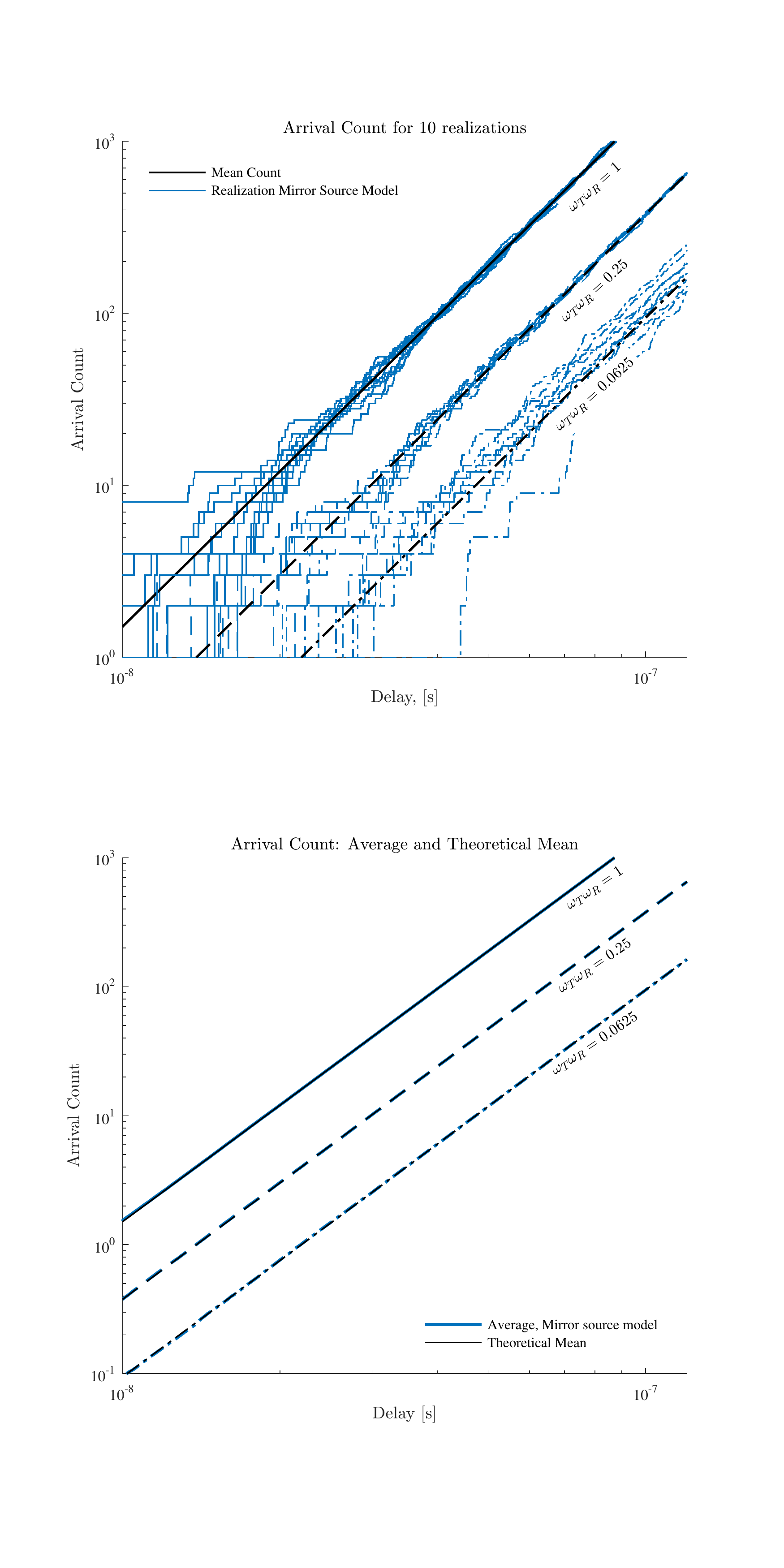}
  \caption{Realizations of arrival count for random transmitter and
    receiver position and orientation (upper panel) and mean arrival
    count (lower panel). }
\label{fig:arrivalRate}
\end{figure}

Fig.~\ref{fig:arrivalRate} reports individual realizations and mean
arrival counts for three different antenna settings. From the realizations depicted in the
upper panel it is evident that the arrival count varies between
realizations and that this variation is more pronounced for more
directive antennas. Moreover, the realizations tend to the mean value
at large delays. The lower panel compares the theoretical expressions
for the mean count \eqref{eq:75} with the Monte Carlo estimates. As
expected, the corresponding curves fit almost perfectly.

\begin{figure}
  \centering
\includegraphics[width=\linewidth]{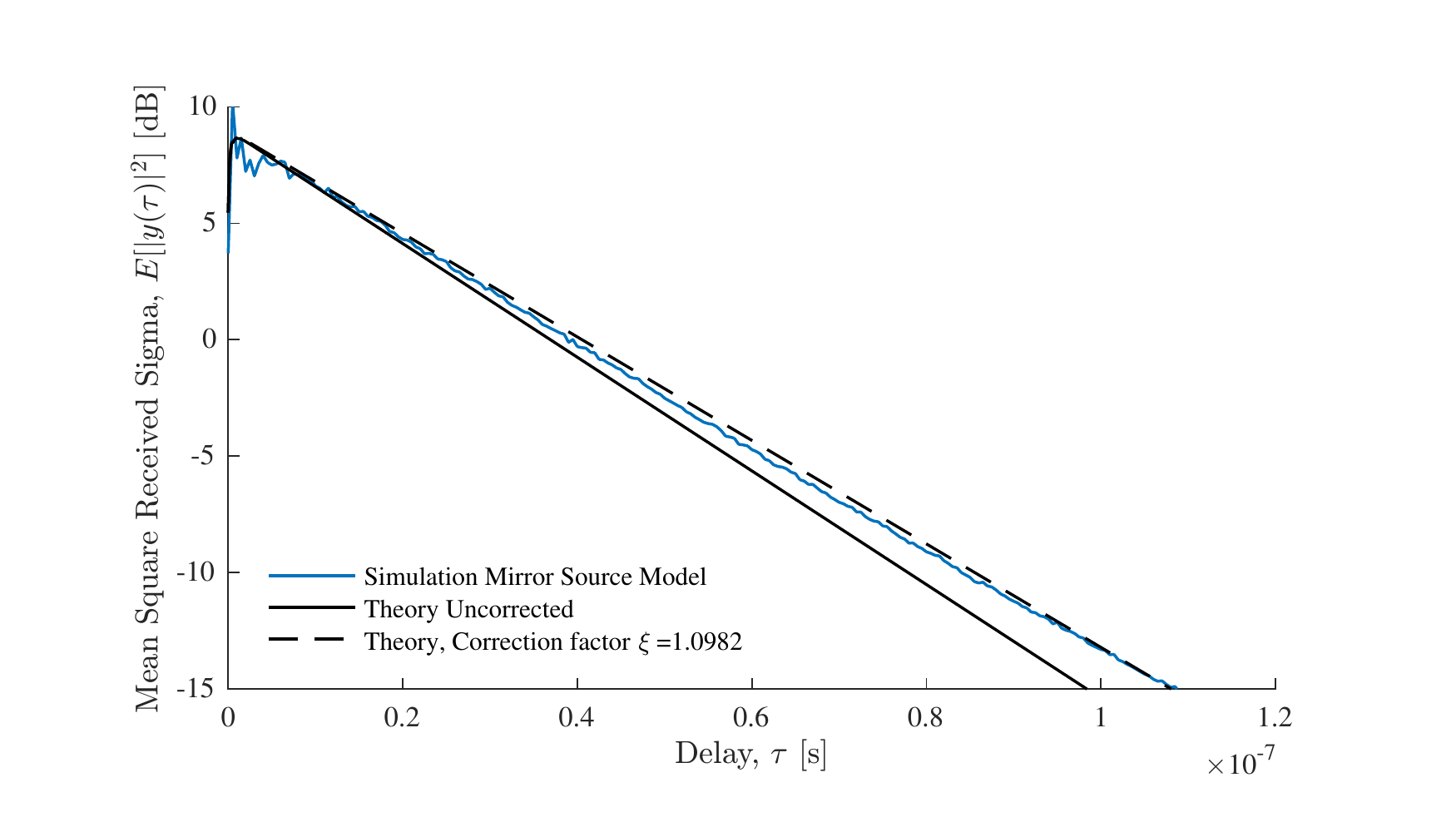}
\caption{Mean square received signal.}
  \label{fig:pds}
\end{figure}

Fig.~\ref{fig:pds} shows the expected power of the received signal,
i.e. $\mathbb E[|y(t)|^2]$, computed using the Monte Carlo
simulation. For clarity, only the results for isotropic antennas are
shown; the curves for the non-isotropic antennas are identical modulo
uncertainties due to the Monte Carlo simulation technique.  This
observation confirms the observation made in the introduction that
models with very different arrival rates, e.g. due to differences in
antenna directivity, can indeed lead to the same power-delay spectrum.
The simulation is compared with the approximation obtained by using
\eqref{eq:93} and \eqref{eq:2}. From Fig.~\ref{fig:pds} it appears
that the slope of the theoretical curve, i.e. the reverberation time
computed in \eqref{eq:103}, deviates from the simulation by about
9\,\%.  The fit can be improved by applying the correction factor
\eqref{eq:1}. According to \cite{Kuttruff2000}, the
value $\gamma^2 = 0.35$ can be used for the aspect ratio of the room considered. For our
simulation setup, this yields a correction factor of $\xi\approx
1.0982$  which gives an excellent fit.

\begin{figure}
  \centering

\includegraphics[width=\linewidth]{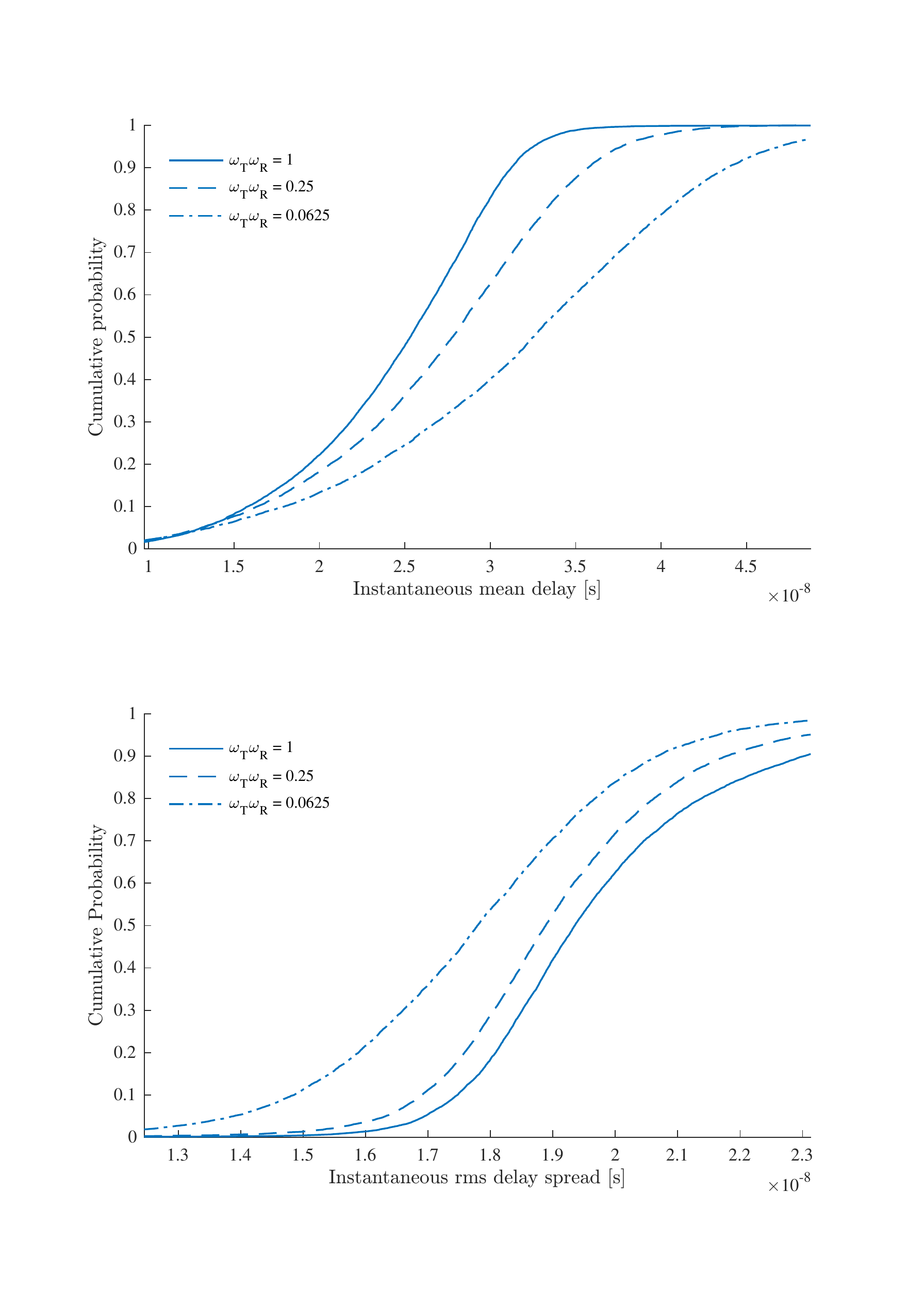}
  \caption{Empirical cumulative probability of  instantaneous mean
    delay (upper panel) and  rms delay spread (lower panel)  with $\omega_T\omega_R$ as parameter. }
  \label{fig:rmsDelay}
\end{figure}

The instantaneous mean delay and rms delay spread are often considered
as important parameters for design of radio systems.  Theoretical
analysis of these entities is beyond the scope of this contribution,
but we report some simulated empirical cumulative distribution
functions in Fig.~\ref{fig:rmsDelay} with $\omega_T\omega_R$ as
parameter. To limit the computational complexity, only 
components with a delay less that 120\,\nano\second\ are simulated.  
In these simulations the mean delay and rms delay spread
are computed as respectively the first and centered second moments of
the realizations of  $|y(t)|^2$ (thus including the effect of
the transmitted pulse).  Even though the antenna directivity does not
affect the  power delay spectrum, it  is apparent
from  Fig.~\ref{fig:rmsDelay} that the instantaneous mean delay and
rms delay spread vary significantly with the antenna directivity. 



\section{Conclusion}
\label{sec:conclusion}
The present study shows how the path arrival rate can be analyzed based on mirror
source theory for rectangular room.  Directivity of the
transmitter and receiver antennas is accounted for by incorporating a simplified
antenna model.  For this setup we give an exact formula for the mean arrival 
count and consequently for the arrival rate for the case  where the
position and orientation of the transmitter are 
uniformly random.  The rate grows quadratically with delay giving rise to a transition
from early  isolated signal components  gradually merging into a
diffuse reverberation tail at later delays. The rate is inversely
proportional to the room  volume, and thus larger rooms lead to a
slower transition. Moreover, the rate is proportional to the product of beam
coverage fractions of the transmitter and receiver antennas, and thus
more directive antennas yield lower arrival rate.  The
derived expression quantifies the impact of directive antennas on the
arrival rate, a phenomenon observed qualitatively in a number of
previous experimental and simulation studies in the literature.

We present  two immediate application of the expression of the arrival rate. First,
we derive a simple formula  for the ``mixing time'', i.e.  the
point in time at which the mean arrival rate exceeds one component per
transmit pulse duration.   The mixing time quantifies to what extent non-overlapping signal
components  is to be expected for a given scenario. 
Second, we use our expression to approximate the power delay
spectrum which appears to be  unaffected by the antenna radiation
pattern. However, the antennas do indeed play an 
important role for the distribution of instantaneous mean delay and 
rms delay spread as shown by Monte Carlo simulations.

The motivation for this work was to achieve calibration of the arrival
rate used in stochastic radio channel models based on geometric
considerations rather than empirically. Indeed this method seems to be
feasible for obtaining models of simplified structures such as the
rectangular room considered here.  The results obtained in the
simplified settings may be used  as building blocks for constructing 
more  for more complex radio propagation environments.



  
\appendices

\section{Second Moment of Arrival Count}
\label{sec:appr-raw-second}
The  raw second moment of the arrival count reads
\begin{align}
\mathbb E[N(\tau)^2] = &\sum_{k,k'}  \mathbb E[N_k  N_{k'}], 
  \label{eq:76}
\end{align}
with the shorthand notation
\begin{equation}
  \label{eq:32}
 N_k = \mathds 1 (\tau<\tau_k) \mathds 1( \Omega_{Tk'}\in \Omega_T )   \mathds 1( \Omega_{Rk'}\in\Omega_R )] .
\end{equation}
Noting that $N_k^2 = N_k$, we see that the sum of diagonal
terms ($k=k'$) equals the mean $\mathbb E[N(\tau)]$ and thus 
\begin{align}
\mathbb E[N(\tau)^2] = & \mathbb E [N(\tau)] + \sum_{k\neq k'}  \mathbb E[N_k  N_{k'}] .
\end{align}
The cross terms $(k\neq k')$, cannot be readily computed. Instead, we approximate
the cross terms by considering the positions of the mirror to be
uncorrelated:
\begin{align}
\mathbb E[N(\tau)^2] \approx & \mathbb E [N(\tau)] +  \sum_{k ,k'}
                               \mathbb E[N_k] \mathbb E[  N_{k'}]  \\
=& \mathbb E [N(\tau)]^2    +\sum_{k} ( \mathbb E[N_k ]  - \mathbb E[N_k ]^2 ).
\end{align}
The terms in the last sum are variances of $N_k$ of which most
vanish.  Only mirror rooms which can be intersected by a
sphere of radius $c\tau$ centered at the receiver contribute to this
sum.  Considering a receiver at the center of the room, for these mirror rooms, 
\begin{equation}
  \label{eq:44}
\tau-D/2c <\tau_k<\tau + D/2c  
\end{equation}
where $D = \sqrt{L_x^2+L_y^2+L_z^2}$ is the length of the main
diagonal of the room. The number of such mirror rooms can be
approximated as 
\begin{equation}
  \label{eq:61}
\mathbb E[N(\tau+D/(2c))]-\mathbb E[N(\tau-D/(2c))]
\end{equation}
Finally, approximating the values of the variances by 
the maximal variance of a Bernoulli variable, we have
\begin{align}
  \label{eq:71}
\mathbb E[N(\tau)^2] \approx& \mathbb E [N(\tau)]^2 \\
& +  \frac{1}{4} (\mathbb E[N(\tau+\frac{D}{2c})]-\mathbb E[N(\tau-\frac{D}{2c})])
\end{align}
Monte Carlo simulations (not reported here) for the setup described in 
Section~\ref{sec:simul-stud-exampl} demonstrate that the approximation
is reasonably accurate for the raw moment, but overshoots the
variance significantly.




 \section{Transmitter with Random Position and
   Fixed Orientation}
\label{sec:mean-arrival-count}
Let the transmitter's  orientation be fixed, but its position
be uniformly distributed.The position and orientation of
the receiver is fixed.  Then the mean arrival count reads 
\begin{align}
  \mathbb E[N(\tau)] &=\mathbb E\big[ 
 \sum_{k}  
\mathds  1(\tau_k<\tau  ) \cdot 
\mathds 1( \Omega_{Tk}\in \Omega_T ) \cdot 
\mathds 1( \Omega_{Rk}\in\Omega_R )
\big]\notag \\
& \leq \mathbb E\big[ 
 \sum_{k}  
\mathds  1(\tau_k<\tau  ) \cdot 
\mathds 1( \Omega_{Rk}\in\Omega_R )\big]
  \label{eq:55}
\end{align}
with equality for isotropic  transmitter antenna. By  Campbell's
theorem,  
\begin{align}
\nonumber  \mathbb E[N(\tau)] &\leq 
  \frac{1}{V}\int  
\mathds 1\left (\tfrac{\|r- r_R\|}{c}<\tau \right) 
\mathds 1\left (\tfrac{r-r_R}{\|r- r_R\|}\in\Omega_R \right) 
 d r  
\\&= \frac{4\pi c^3\tau^3 }{3V}\cdot  \omega_R \mathds 1 (\tau>0).
  \label{eq:62}
\end{align}
Symmetry gives a similar inequality involving $\omega_R$.  In
combination, these two lower bound yields
\begin{align}
  \label{eq:65}
  \mathbb E[N(\tau)] &\leq \frac{4\pi c^3\tau^3 }{3V}\cdot  \min \{ \omega_T,\omega_R\}\mathds 1 (\tau>0),
\end{align}
again, with equality obtained either of the antennas are
isotropic. Since \eqref{eq:65} holds for all $\tau$, the arrival rate
is upper bounded as
\begin{equation}
  \label{eq:81}
  \lambda(\tau)  \leq \frac{4\pi c^3\tau^2 }{V}\cdot  \min \{ \omega_T,\omega_R\}\mathds 1 (\tau>0).
\end{equation}
with equality if either of the antennas are isotropic.

We remark that by symmetry, the bounds \eqref{eq:65} and \eqref{eq:81} hold true
if we instead let position of the receiver be uniformly distributed
within the room and the transmitters be fixed. Furthermore, it can be
shown by some adaptation of the proof that the bound also holds in the
case where both transmitter and receiver have independent and
uniformly distributed but fixed orientations.

\section{Deterministic Transmitter-Receiver Distance}
\label{sec:determ-transm-rece}
To compute the mean arrival count for fixed
transmitter-receiver distance we need to compute a conditional
expectation. However, the condition renders the calculation of the mean count
very cumbersome if at all possible.  Instead, 
we approximate the expected count  as motivated by the following reasoning. First, the conditional arrival count is
strictly zero for $\tau<\tau_0$. Second, due to the random orientation
of antennas, the direct component $\tau = \tau_0$ occurs with
probability $\omega_T\omega_R$. Third,
conditioning on $\tau_0$ does not change the fact that there is
exactly one mirror source per mirror room.  Therefore, the
mean count for $c\tau$ much greater than the diagonal of the room
remains the same as in the unconditional case. Thus,  we have the
approximation for the conditional  mean arrival count
\begin{equation}
  \label{eq:82}
 \mathbb E[N(\tau)|\tau_0] \approx \mathds 1 (\tau \leq \tau_0) \left(1
 + \frac{4\pi c^3(\tau^3-\tau_0^3)}{3V}\right) \omega_T \omega_R 
\end{equation}
with corresponding conditional arrival rate 
\begin{equation}
\label{eq:83}
\lambda(\tau|\tau_0) \approx\delta(\tau-\tau_0) \omega_T \omega_R + 
  \mathds 1(\tau>\tau_0)\frac{4\pi c^3\tau^2}{V} \omega_T \omega_R. 
\end{equation}
The right hand side of \eqref{eq:82} coincides with 
that of the approximation obtained in the case with non-random 
transmitter and receiver location in \eqref{eq:53}.

\vspace{-1ex}

\bibliographystyle{IEEEtran}
\bibliography{../../referenceDataBase/referencedatabase}

\begin{thebibliography}{10}
\providecommand{\url}[1]{#1}
\csname url@samestyle\endcsname
\providecommand{\newblock}{\relax}
\providecommand{\bibinfo}[2]{#2}
\providecommand{\BIBentrySTDinterwordspacing}{\spaceskip=0pt\relax}
\providecommand{\BIBentryALTinterwordstretchfactor}{4}
\providecommand{\BIBentryALTinterwordspacing}{\spaceskip=\fontdimen2\font plus
\BIBentryALTinterwordstretchfactor\fontdimen3\font minus
  \fontdimen4\font\relax}
\providecommand{\BIBforeignlanguage}[2]{{%
\expandafter\ifx\csname l@#1\endcsname\relax
\typeout{** WARNING: IEEEtran.bst: No hyphenation pattern has been}%
\typeout{** loaded for the language `#1'. Using the pattern for}%
\typeout{** the default language instead.}%
\else
\language=\csname l@#1\endcsname
\fi
#2}}
\providecommand{\BIBdecl}{\relax}
\BIBdecl

\bibitem{Hashemi1993}
H.~Hashemi, ``The indoor radio propagation channel,'' \emph{Proc. {IEEE}},
  vol.~81, no.~7, pp. 943--968, Jul. 1993.

\bibitem{turin}
G.~Turin, F.~Clapp, T.~Johnston, S.~Fine, and D.~Lavry, ``A statistical model
  of urban multipath propagation channel,'' \emph{{IEEE} Trans. Veh. Technol.},
  vol.~21, pp. 1--9, Feb. 1972.

\bibitem{suzuki}
H.~Suzuki, ``A statistical model for urban radio propagtion channel,''
  \emph{{IEEE} Trans. on Commun. Syst.}, vol.~25, pp. 673--680, Jul. 1977.

\bibitem{hashemi}
H.~Hashemi, ``Simulation of the urban radio propagation,'' \emph{{IEEE} Trans.
  Veh. Technol.}, vol.~28, pp. 213--225, Aug. 1979.

\bibitem{saleh}
A.~A.~M. Saleh and R.~A. Valenzuela, ``A statistical model for indoor multipath
  propagation channel,'' \emph{{IEEE} J. Sel. Areas Commun.}, vol. SAC-5,
  no.~2, pp. 128--137, Feb. 1987.

\bibitem{Spencer2000}
Q.~H. Spencer, B.~Jeffs, M.~Jensen, and A.~Swindlehurst, ``Modeling the
  statistical time and angle of arrival characteristics of an indoor multipath
  channel,'' \emph{{IEEE} J. Sel. Areas Commun.}, vol.~18, no.~3, pp. 347--360,
  2000.

\bibitem{Zwick2002}
T.~Zwick, C.~Fischer, and W.~Wiesbeck, ``A stochastic multipath channel model
  including path directions for indoor environments,'' \emph{{IEEE} J. Sel.
  Areas Commun.}, vol.~20, no.~6, pp. 1178--1192, Aug. 2002.

\bibitem{Zwick2000}
T.~Zwick, C.~Fischer, D.~Didascalou, and W.~Wiesbeck, ``A stochastic spatial
  channel model based on wave-propagation modeling,'' \emph{{IEEE} J. Sel.
  Areas Commun.}, vol.~18, no.~1, pp. 6--15, Jan. 2000.

\bibitem{Haneda2015}
\BIBentryALTinterwordspacing
K.~Haneda, J.~Jarvelainen, A.~Karttunen, M.~Kyro, and J.~Putkonen, ``A
  statistical spatio-temporal radio channel model for large indoor environments
  at 60 and 70 {GHz},'' \emph{{IEEE} Transactions on Antennas and Propagation},
  vol.~63, no.~6, pp. 2694--2704, jun 2015. [Online]. Available:
  \url{http://dx.doi.org/10.1109/tap.2015.2412147}
\BIBentrySTDinterwordspacing

\bibitem{Samimi2016}
\BIBentryALTinterwordspacing
M.~K. Samimi and T.~S. Rappaport, ``{3-D} millimeter-wave statistical channel
  model for 5g wireless system design,'' \emph{{IEEE} Transactions on Microwave
  Theory and Techniques}, vol.~64, no.~7, pp. 2207--2225, jul 2016. [Online].
  Available: \url{http://dx.doi.org/10.1109/TMTT.2016.2574851}
\BIBentrySTDinterwordspacing

\bibitem{Holloway1999}
C.~Holloway, M.~Cotton, and P.~McKenna, ``A model for predicting the power
  delay profile characteristics inside a room,'' \emph{{IEEE} Trans. Veh.
  Technol.}, vol.~48, no.~4, pp. 1110--1120, July 1999.

\bibitem{Rudd2003}
R.~Rudd and S.~Saunders, ``Statistical modelling of the indoor radio channel --
  an acoustic analogy,'' in \emph{Proc. Twelfth International Conf. on Antennas
  and Propagation (Conf. Publ. No. 491)}, vol.~1, 31 March--3 April 2003, pp.
  220--224.

\bibitem{Rudd2007}
R.~F. Rudd, ``The prediction of indoor radio channel impulse response,'' in
  \emph{The Second European Conf. on Antennas and Propagation, 2007. EuCAP
  2007.}, Nov. 2007, pp. 1--4.

\bibitem{Andersen2007}
J.~B. Andersen, J.~{\O}. Nielsen, G.~F. Pedersen, G.~Bauch, and J.~M. Herdin,
  ``Room electromagnetics,'' \emph{{IEEE} Antennas Propag. Mag.}, vol.~49,
  no.~2, pp. 27--33, Apr. 2007.

\bibitem{Bamba2012}
\BIBentryALTinterwordspacing
A.~Bamba, W.~Joseph, J.~B. Andersen, E.~Tanghe, G.~Vermeeren, D.~Plets,
  J.~{\O}. Nielsen, and L.~Martens, ``Experimental assessment of specific
  absorption rate using room electromagnetics,'' \emph{{IEEE} Transactions on
  Electromagnetic Compatibility}, vol.~54, no.~4, pp. 747--757, aug 2012.
  [Online]. Available: \url{http://dx.doi.org/10.1109/TEMC.2012.2189572}
\BIBentrySTDinterwordspacing

\bibitem{Steinbock2015}
\BIBentryALTinterwordspacing
G.~Steinboeck, T.~Pedersen, B.~H. Fleury, W.~Wang, and R.~Raulefs,
  ``Experimental validation of the reverberation effect in room
  electromagnetics,'' \emph{{IEEE} Trans. Antennas Propagat.}, vol.~63, no.~5,
  pp. 2041--2053, may 2015. [Online]. Available:
  \url{http://dx.doi.org/10.1109/TAP.2015.2423636}
\BIBentrySTDinterwordspacing

\bibitem{Kuttruff2000}
H.~Kuttruff, \emph{Room Acoustics}.\hskip 1em plus 0.5em minus 0.4em\relax
  London: Taylor {\&} Francis, 2000.

\bibitem{Steinboeck2013d}
G.~Steinb\"{o}ck, T.~Pedersen, B.~Fleury, W.~Wang, and R.~Raulefs,
  ``{Calibration of the Propagation Graph Model in Reverberant Rooms},'' in
  \emph{{URSI Commission F Triennial Open Symposium on Radiowave Propagation
  and Remote Sensing}}, May 2013.

\bibitem{Steinboeck2016}
\BIBentryALTinterwordspacing
G.~Steinboeck, M.~Gan, P.~Meissner, E.~Leitinger, K.~Witrisal, T.~Zemen, and
  T.~Pedersen, ``Hybrid model for reverberant indoor radio channels using rays
  and graphs,'' \emph{{IEEE} Transactions on Antennas and Propagation},
  vol.~64, no.~9, pp. 4036--4048, sep 2016. [Online]. Available:
  \url{http://dx.doi.org/10.1109/tap.2016.2589958}
\BIBentrySTDinterwordspacing

\bibitem{Steinboeck2013}
\BIBentryALTinterwordspacing
G.~Steinbock, T.~Pedersen, B.~H. Fleury, W.~Wang, and R.~Raulefs, ``Distance
  dependent model for the delay power spectrum of in-room radio channels,''
  \emph{{IEEE} Trans. Antennas Propag.}, vol.~61, no.~8, pp. 4327--4340, aug
  2013. [Online]. Available: \url{http://dx.doi.org/10.1109/tap.2013.2260513}
\BIBentrySTDinterwordspacing

\bibitem{Jakobsen2012}
\BIBentryALTinterwordspacing
M.~L. Jakobsen, B.~H. Fleury, and T.~Pedersen, ``Analysis of the stochastic
  channel model by saleh \&amp; valenzuela via the theory of point processes,''
  in \emph{Int. Zurich Seminar on Communications (IZS), February 29 - March 2,
  2012}.\hskip 1em plus 0.5em minus 0.4em\relax Z{\"u}rich, Eidgen{\"o}ssische
  Technische Hochschule Z{\"u}rich, 2012. [Online]. Available:
  \url{https://doi.org/10.3929/ethz-a-007052489}
\BIBentrySTDinterwordspacing

\bibitem{Eyring1930}
C.~F. Eyring, ``Reverberation time in 'dead' rooms,'' \emph{The Journal of the
  Acoustical Society of Amarica}, vol.~1, no.~2, p. 241, 1930.

\bibitem{Kunisch2003a}
J.~Kunisch and J.~Pamp, ``{UWB} radio channel modeling considerations,'' in
  \emph{Proc. International Conference on Electromagnetics in Advanced
  Applications 2003}, Turin, Sep. 2003.

\bibitem{Kunisch2002}
------, ``Measurement results and modeling aspects for the {UWB} radio
  channel,'' in \emph{IEEE Conf. on Ultra Wideband Systems and Technologies,
  2002. Digest of Papers}, May 2002, pp. 19--24.

\bibitem{Pedersen2012}
T.~Pedersen, G.~Steinb{\"o}ck, and B.~H. Fleury, ``Modeling of reverberant
  radio channels using propagation graphs,'' \emph{{IEEE} Trans. Antennas
  Propag.}, vol.~60, no.~12, pp. 5978--5988, Dec. 2012.

\bibitem{Pedersen2007}
T.~Pedersen and B.~Fleury, ``Radio channel modelling using stochastic
  propagation graphs,'' in \emph{Proc. IEEE International Conf. on
  Communications ICC '07}, Jun. 2007, pp. 2733--2738.

\bibitem{Manabe1996}
T.~Manabe, Y.~Miura, and T.~Ihara, ``Effects of antenna directivity and
  polarization on indoor multipath propagation characteristics at 60 ghz,''
  \emph{{IEEE} J. Sel. Areas Commun.}, vol.~14, no.~3, pp. 441--448, apr 1996.

\bibitem{Goodman2006}
N.~A. Goodman and K.~L. Melde, ``The impact of antenna directivity on the
  small-scale fading in indoor environments,'' \emph{{IEEE} Trans. Antennas
  Propag.}, vol.~54, no.~12, pp. 3771--3777, Dec. 2006.

\bibitem{Yang2008}
\BIBentryALTinterwordspacing
H.~Yang, M.~Herben, I.~Akkermans, and P.~Smulders, ``Impact analysis of
  directional antennas and multiantenna beamformers on radio transmission,''
  \emph{{IEEE} Transactions on Vehicular Technology}, vol.~57, no.~3, pp.
  1695--1707, may 2008. [Online]. Available:
  \url{https://doi.org/10.1109%2Ftvt.2007.907308}
\BIBentrySTDinterwordspacing

\bibitem{Smulders2009}
\BIBentryALTinterwordspacing
P.~Smulders, ``Statistical characterization of 60-{GHz} indoor radio
  channels,'' \emph{{IEEE} Transactions on Antennas and Propagation}, vol.~57,
  no.~10, pp. 2820--2829, oct 2009. [Online]. Available:
  \url{https://doi.org/10.1109%2Ftap.2009.2030524}
\BIBentrySTDinterwordspacing

\bibitem{Friis1946}
H.~T. Friis, ``A note on a simple transmission formula,'' \emph{Proceedings of
  the I.R.E.}, vol.~34, no.~5, pp. 254--256, may 1946.

\bibitem{Neubauer2001}
R.~Neubauer and B.~Kostek, ``Prediction of the reverberation time in
  rectangular rooms with non-uniformly distributed sound absorption,''
  \emph{Archives of Acoustics}, vol.~26, no.~3, pp. 183--201, 2001.

\bibitem{Stoyan1995}
D.~Stoyan, W.~S. Kendall, and J.~Mecke, \emph{Stochastic Geometry and its
  Applications}, second edition~ed.\hskip 1em plus 0.5em minus 0.4em\relax John
  Wiley \& Sons, Inc., 1995.

\bibitem{Lindau2010}
A.~Lindau, L.~Kosanke, and S.~Weinzierl, ``Perceptual evaluation of physical
  predictors of the mixing time in binaural room impulse responses,'' in
  \emph{Audio Engineering Society Convention 128}.\hskip 1em plus 0.5em minus
  0.4em\relax Audio Engineering Society, 2010.

\end{thebibliography}


\end{document}